\documentclass{article}

\usepackage[style=alphabetic,backend=bibtex]{biblatex}
\bibliography{references}

\usepackage{graphicx}
\usepackage{tikz}
\usepackage{amsmath}
\usepackage{hyperref}
\usepackage{amssymb}
\usepackage{url}

\usepackage{bm}

\usepackage{tikz}
\usetikzlibrary{arrows,automata,patterns,patterns.meta}

\tikzset{
  treenode/.style = {align=center, inner sep=0pt, text centered,
    font=\sffamily},
  arn_n/.style = {treenode, circle, white, font=\sffamily\bfseries, draw=black,
    fill=black, text width=1.5em},
  arn_r/.style = {treenode, circle, red, draw=red, 
    text width=1.5em, very thick},
  arn_x/.style = {treenode, rectangle, draw=black,
    minimum width=0.5em, minimum height=0.5em}
}

\newcommand{\reals}{{\mbox{\bf R}}}

\newcommand{\naturals}{{\mbox{\bf N}}}

\newcommand{\Expect}{\mathop{\bf E{}}}

\newcommand{\Prob}{\mathop{\bf Prob}}

\newcommand{\eg}{{\it e.g.}}
\newcommand{\ie}{{\it i.e.}}

\newcommand{\BEAS}{\begin{eqnarray*}}
\newcommand{\EEAS}{\end{eqnarray*}}
\newcommand{\BEA}{\begin{eqnarray}}
\newcommand{\EEA}{\end{eqnarray}}
\newcommand{\BEQ}{\begin{equation}}
\newcommand{\EEQ}{\end{equation}}
\newcommand{\BIT}{\begin{itemize}}
\newcommand{\EIT}{\end{itemize}}

\begin{document}
\title{Attacks on Dynamic DeFi Interest Rate Curves}
\author{Tarun Chitra, Peteris Erins, Kshitij Kulkarni}
%
\maketitle              
\begin{abstract}
As decentralized money market protocols continue to grow in value locked, there have been a number of optimizations proposed for improving capital efficiency. One set of proposals from Euler Finance and Mars Protocol is to have an interest rate curve that is a proportional-integral-derivative (PID) controller. In this paper, we demonstrate attacks on proportional and proportional-integral controlled interest rate curves. The attack allows one to manipulate the interest rate curve to take a higher proportion of the earned yield than their pro-rata share of the lending pool. We conclude with an argument that PID interest rate curves can actually \emph{reduce} capital efficiency (due to attack mitigations) unless supply and demand elasticity to rate changes are sufficiently high.
\end{abstract}

\section{Introduction}
As decentralized finance, or DeFi, has grown since 2018, one of the top use cases is on-chain, overcollateralized lending.
These loans involve three agents --- lenders, borrowers, and liquidators --- whose capital is coordinated via sets of smart contracts.
Lenders pool together assets that they want to earn yield on within a smart contract that pools their funds and makes them available to borrowers.
Borrowers deposit collateral of one type of asset and borrow another type of asset, with the maximum quantity they can borrow controlled by a governance controlled parameter referred to as either loan-to-value (LTV) or collateral factor (CF).
Finally, liquidators buy defaulted borrower collateral from the smart contract when a borrower's borrowed liabilities are worth more than the assets they've placed as collateral. \\ \\
\noindent Lending protocols on Ethereum, Binance Smart Chain, Polygon, Tron, Avalanche, Arbitrum, Optimism and Solana currently have \$14 billion of assets on them~\cite{0xngmi_2022}.
Most of these funds are assets deposited by lenders who earn yield from borrowers (whose collateral is also included within the \$14 billion calculation).
In order to have orderly operation of these protocols, liquidators and/or pooled liquidation mechanisms are needed to ensure that bad debt is cleared from the system.
The parameters set by the protocol, such as LTV and liquidation incentives, control how efficiently the protocol is matching borrow demand with pooled supply.

\paragraph{Capital Efficiency for Overcollateralized Loans}
Unlike traditional Lombard-style overcollateralized loans, on-chain lenders such as Aave and Compound currently do not price discriminate against borrowers.
This is because DeFi protocols can be used pseudonymously and there is no incontrovertible form of sybil resistant identity to use for credit scoring.
As such, these protocols have to be significantly less capital efficient than their centralized counterparts to ensure that bad debt can be liquidated successfully.
Formally, capital efficiency can be defined as the maximal ratio of the amount of revenue generating capital to the quantity of capital reserves needed to facilitate risky operations.
The denominator of this quantity is necessarily higher in decentralized systems in order to avoid cascading liquidations (see, \eg, Aave liquidation analysis~\cite{fuAave}). \\ \\ 
\noindent The main mechanisms for improving capital efficiency in DeFi involve using dynamic parameter updates to risk parameters as a function of market conditions.
For instance, if liquidity increases for various collateral assets, LTVs can be increased generating more revenue for the protocol and increasing capital efficiency.
On the other hand, when tail events occur, such as the stETH/ETH deviation from par incident in May 2022, these parameters can be reduced to decrease the likelihood of cascading liquidations.
Aave, Compound, and MakerDAO have had dynamic risk parameter updates to improve the risk and capital efficiency trade-off from a variety of DAO participants~\cite{morrow_2022}. \\ \\
\noindent However, few protocols have had dynamic interest rate curve optimizations.
In theory, dynamic interest rates could optimize capital efficiency in a manner that is equivalent to credit scoring.
A dynamic interest rate can adjust with regards to more than simply the utilization (\eg~ratio of demanded borrow to supplied assets in num\'eraire terms).
For instance, a dynamic rate could adjust in response to particular borrowers (which is analogous to credit scoring) or in response to a coherent risk measure (such as a moving average of liquidations or insolvencies). \\ \\ 
\noindent Practically speaking, however, it is extremely difficult to implement complex dynamic interest rate mechanisms on-chain.
The computational and storage limitations of blockchains make it difficult to implement on-chain logic for complex scoring and/or rating of users.
While there have been a number of protocols have sprung up to do off-chain credit scoring and/or rate updates~\cite{wohrle_2021, cred}, these protocols have not proven to be able to attract large quantities of liquidity.
\paragraph{On-Chain Interest Rate Controllers.}
There has also been a focus on on-chain implementable dynamic interest rates using simple mechanisms such as proportional-integral-derivative (PID) controllers.
Part of the interest in using such methods is that they are both easy to implement and computationally cheap, making a security audit an implementation of a PID controller relatively easy. 
Moreover, protocols such as Reflexer~\cite{blockscience_2021} already use PID-like mechanisms in production on Ethereum.\footnote{We note that Reflexer's implementation of a PID controller is technically closer to a bang-bang controller and/or model predictive control since it has sharp non-smooth points in how it computes its rate.}
The argument made for using these mechanisms for on-chain lending is that PID controllers allow for the risk-reward trade-off to be tuned dynamically as a function of time-weighted averages of the supply, demand, and utilization of the protocol. \\ \\
\noindent The two major designs proposed for PID controlled interest rate curves are Euler Finance's reactive rates~\cite{bentley_2022, bentley_hoyte_2022} and Mars Protocol.
Currently, only Mars Protocol has implemented~\cite{mars_github, mars_medium} and deployed a proportional interest rate controller to production.
Euler Finance has signalled that they would introduce reactive rates pending further research.
However, it appears that neither of these teams has formally analyzed their PID interest protocols. \\ \\
\noindent We note that most PID designs in DeFi are either proportional (P) or proportional-integral (PI) controllers.
The PI controllers correspond to the utilization of time-weighted average quantities (akin to the Uniswap V3 TWAP oracle~\cite{mackinga2022twap}).
There are two reasons derivatives of rate changes are less useful in practice.
First, the rate of change of an interest rate is more easily manipulable given the constraints of blockchains, such as large confirmation times.
Moreover, the only reason to adjust a rate based on its gradient is to provide fixed-interest rate protocols.
However, most fixed-interest rate protocols such as Yield and Notional, use more transaction cost efficient mechanisms than a PID controller~\cite{yield}.

\paragraph{This Paper.}
To construct a formal analysis of PID interest rate controllers, we present a simple attack on both proportional (P) and proportional-integral (PI) controllers.
Virtually all known controllers used in DeFi are proportional controllers with constraints on the parameters involved. 
The attack we present in \S\ref{sec:proportional} is the lending analogue of a Just-in-Time liquidity attack against an automated market maker such as Uniswap V3~\cite{loesch2021impermanent}.
The high-level idea behind the attack is as follows:
\begin{enumerate}
    \item A strategic user adds a large amount of demand and raises utilization rates by opening borrow (\ie~creating loans), causing the interest rate to spike dramatically.
    \item The choice of how much of a spike was caused allows the strategic user to sequentially reduce demand (repaying a portion of the opened loans) and increase supply in such a way that the interest rate and utilization are stationary.
    \item After a number of blocks, the strategic user returns the utilization to the level prior to opening loans.
\end{enumerate}
The profit earned by the strategic user comes at the expense of existing supplying lenders, who effectively realize less interest payments than they should have earned. 

The profitability of this strategy depends on whether other lenders or borrowers change the utilization rate and on the quantity of assets demanded and supplied.
We analyze profitability in the worst case for lenders: when there is no supply or demand elasticity.
The supply and/or demand elasticity of a protocol refers to the expected rate of change of supply or demand in the protocol as a function of a rate change.
We usually have the supply elasticity be positive when rates increase whereas the demand elasticity is negative.
However, in many DeFi protocols there are a large swath of users who are completely inelastic to rate changes~\cite{dc_2022}.
This means that the worst case condition of the attack is often true. \\ \\ 
\noindent We also analyze how this attack is related to capital efficiency in the protocol.
We demonstrate that the attack has low profitability if there is excess capital within the pool (\eg~the utilization rate and the target utilization rate are low and the supplied assets dwarf the demand).
Similarly, if the protocol can time-lock assets (\eg~force a user who to supply or borrow for a minimum time period), the protocol can make such an attack significantly more expensive
However, both of these options are very capital inefficient states for a lending protocol. \\ \\
\noindent Our results show that in the case of manipulating the proportional (P) interest rate curve, the strategic user's profit can be decomposed into two components: one depending on the duration of the attack and one on a `hurdle rate' or minimum threshold depending on the current utilization. The profit also depends on the spread between the borrow and supply rates, given by a constant $\gamma \in (0,1]$ and increases with this spread. We further extend this analysis to proportional-integral (PI) interest rate curves, where the profit is reduced by the presence of additional costs that the attacker must pay to first manipulate the supply and demand to desired levels and then extract yield. 
\noindent Therefore, our solution for mitigating attacks on proportional controllers involves three components:
\begin{enumerate}
    \item Using a PI controller which is more expensive to attack (see Appendix \ref{app:PI_profit})
    \item Separating supply and demand curves (akin to what Compound V3 does~\cite{leshner_2022})
    \item Having the controller depend not only on utilization but also supply and demand elasticities
\end{enumerate}
We further note that any off-chain optimization of interest rate curves should take attack profitability into account. 
\section{Dynamically Controlled Interest Rates}\label{sec:proportional}
Proportional controlled interest rates can be griefed by strategic users in a plethora of ways.
The majority of these griefing attacks involve a strategic user modifying the interest rate in a sufficient way that they put up a small amount of initial capital but then extract a higher than pro-rata share of the supply interest.
In this section, we will illustrate a simple griefing attack and provide sufficient conditions, dependent on the probability distributions for borrower demand and lender supply, for such an attack to be profitable.

\paragraph{Model Setup.}
For simplicity, we will assume that the interest rate curve for the protocol is a single linear curve.
The commonly used kinked demand curve is a piecewise linear function, so the same attack naturally generalizes to the piecewise setting.
Other more complex interest rate curves will have varying profitability for this attack, which can be estimated by piecewise linear approximation. \\ \\
\noindent Our system has four dynamic variables that represent the aggregated agent state within the protocol:
\begin{enumerate}
    \item $S_t$: Amount of collateral asset supplied at time $t \in \naturals$
    \item $D_t$: Amount of borrow asset demanded at time $t \in \naturals$
    \item $U_t$: Utilization, defined as $U_t = D_t / S_t$
    \item $k_t(U_t)$: Linear proportionality constant for the rate curve (see eq.~\eqref{eq:prop_const})
\end{enumerate}
The linear borrower interest rate curve $r_t(S_t, D_t)$ is therefore computed as
\[
r_t(S_t, D_t) = \frac{k_t(U_t) D_t}{S_t} = k_t(U_t) U_t
\]
where $U_t = \frac{D_t}{S_t} \in (0, 1)$ is the utilization rate of the protocol.
In most protocols, there is a fixed spread $\gamma \in (0, 1]$, known as the reserve factor in Aave and Compound, that separates the borrow and supply rates.
The supply rate is defined to be $r^S_t(S_t, D_t) = \gamma r_t(S_t, D_t) U_t$. \\ \\ 
\noindent We will focus on proportional rate curves as they are the dominant ones used in both Euler (still not launched) and Mars.\footnote{We note that Mars Protocol is constrained such that for all $t\in\naturals$, $k_t \in [k_-, k_+]\subset \reals_+$ for some $k_-, k_+$~\cite{mars_github}}
Proportional demand rate curves update the linear proportionality constant $k_t(U_t)$ as: 
\begin{equation}\label{eq:prop_const}
k_t(U_t) = k_{t-1}(U_{t-1}) + \alpha (U_t - U^*)
\end{equation}
where $U^*$ is a target utilization rate and $\alpha$ is a governance controlled parameter (akin to the step size in EIP-1559). In the case where the protocol uses a proportional-integral (PI) controller to set rates, the above update is replaced with: 
\begin{equation}\label{eq:prop_integral}
    k_t(U_t) = k_{t-1}(U_{t-1}) + \alpha(U_t - U^{\star}) + \beta \sum_{i = t -p}^{t-1} \xi^i (U_i - U^{\star})
\end{equation}
where $\beta$ is the governance controlled parameter corresponding to the integral component of the controller, $\xi \in (0,1]$ is a decay constant, and $p$ is a constant lookback time.  \\ \\
\noindent Note that in both cases, when utilization at time $t$ is below the target, \ie~$U_t < U^*$, then the interest rate is lowered to attract more borrowers (and vice verse when it is above).
Finally, we assume that no other borrowers or suppliers change the supply during the time period where the strategic user executes the attack.

\begin{figure}
    \centering
    
    \begin{tikzpicture}
      \tikzstyle{every node}=[font=\footnotesize]

      \draw[thick,->] (0,0) -- (7,0) node[anchor=north west] {$t$};
      \draw[thick,->] (0,0) -- (0,2.5) node[anchor=south east] {Utilization, $U_t$};
      
      \draw (1 cm,1pt) -- (1 cm,-1pt) node[anchor=north] {$s-1$};
      \draw (2 cm,1pt) -- (2 cm,-1pt) node[shift={(0,-0.06)}, anchor=north] {$s$};
      \foreach \x in {1,2,3,4}
        \draw (\x cm + 2 cm,1pt) -- (\x cm + 2 cm,-1pt) node[anchor=north] {$s+\x$};
      
      \draw (1pt,0.6 cm) -- (-1pt,0.6 cm) node[anchor=east] {$U_{s-1}$};
      \draw (1pt,1 cm) -- (-1pt,1 cm) node[anchor=east] {$U^*$};
      \draw (1pt,2 cm) -- (-1pt,2 cm) node[anchor=east] {$2U^*$};
      
      \draw (1,0.6) -- (2,0.6) -- (2,2) -- (3,2) -- (3,1) -- (4,1) -- (5,1) -- (6,1);
    \end{tikzpicture}
    
    \begin{tikzpicture}
      \tikzstyle{every node}=[font=\footnotesize]
      
      \hspace*{8.5 pt}

      \draw[thick,->] (0,0) -- (7,0) node[anchor=north west] {$t$};
      \draw[thick,->] (0,0) -- (0,2.5) node[anchor=south east] {Added Supply $\Delta S_t$};
      
      \draw (1 cm,1pt) -- (1 cm,-1pt) node[anchor=north] {$s-1$};
      \draw (2 cm,1pt) -- (2 cm,-1pt) node[shift={(0,-0.06)}, anchor=north] {$s$};
      \foreach \x in {1,2,3,4}
        \draw (\x cm + 2 cm,1pt) -- (\x cm + 2 cm,-1pt) node[anchor=north] {$s+\x$};
      
      \draw (1pt,0.6 cm) -- (-1pt,0.6 cm) node[anchor=east] {$\Delta S_{s-1}  =0$};
      \draw (1pt,2 cm) -- (-1pt,2 cm) node[anchor=east] {$\Delta S_s$};
      
      \draw (1,0.6) -- (2,0.6) -- (2,2) -- (3,2) -- (3,1.9) -- (4,1.9) -- (4,1.8) -- (5,1.8) -- (5,1.7) -- (6,1.7) -- (6,1.6);
    \end{tikzpicture}
    
    \begin{tikzpicture}
      \tikzstyle{every node}=[font=\footnotesize]
      
      \hspace*{6.5 pt}

      \draw[thick,->] (0,0) -- (7,0) node[anchor=north west] {$t$};
      \draw[thick,->] (0,0) -- (0,2.5) node[anchor=south east] {Demand $D_t$};
      
      \draw (1 cm,1pt) -- (1 cm,-1pt) node[anchor=north] {$s-1$};
      \draw (2 cm,1pt) -- (2 cm,-1pt) node[shift={(0,-0.06)}, anchor=north] {$s$};
      \foreach \x in {1,2,3,4}
        \draw (\x cm + 2 cm,1pt) -- (\x cm + 2 cm,-1pt) node[anchor=north] {$s+\x$};
      
      \draw (1pt,0.6 cm) -- (-1pt,0.6 cm) node[anchor=east] {$D_{s-1}$};
      \draw (1pt,1 cm) -- (-1pt,1 cm) node[anchor=east] {$D^*$};
      \draw (1pt,2 cm) -- (-1pt,2 cm) node[anchor=east] {$2D^*$};
      
      \draw[dashed] (0,1) -- (7,1) node[right]{};
      
      \draw (1,0.6) -- (2,0.6) -- (2,2) -- (3,2) -- (3,1.75) -- (4,1.75) -- (4,1.5) -- (5,1.5) -- (5,1.25) -- (6,1.25) -- (6,1);
    \end{tikzpicture}
    
    \caption{Diagram of $U_t, \Delta S_t, D_t$ for the strategic user. The user initially spikes the interest rate at time $s$ and then relaxes it in such a way that the interest rate is kept high even though demand is decreasing (which represents a deadweight loss inured by passive, inelastic suppliers).}
    \label{fig:pid_utilization}
\end{figure}
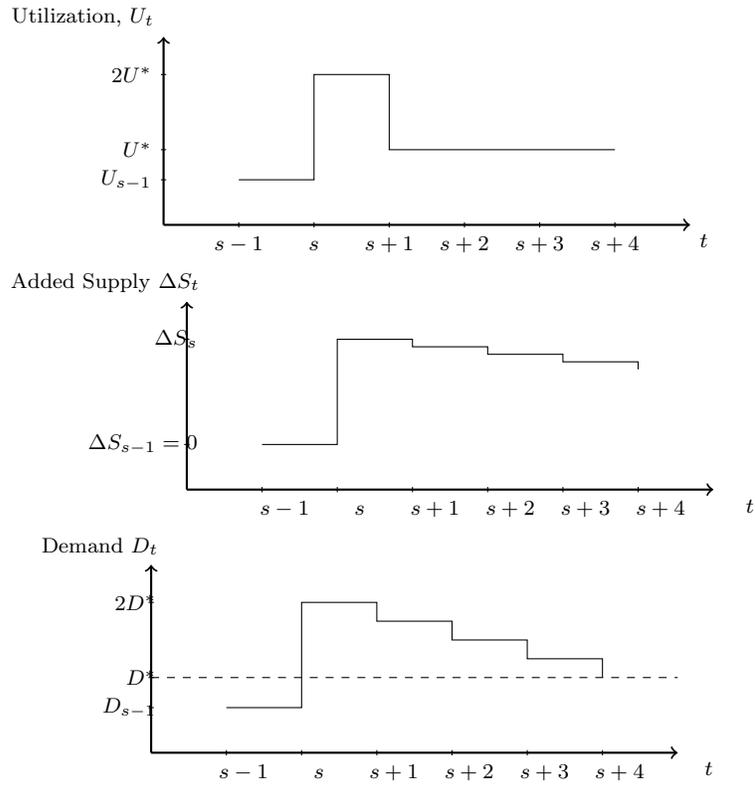

\paragraph{Strategic User Profitability.}
The basic idea for a strategic user's strategy is that they increase the demand at time $s$, causing the interest rate to spike according to \eqref{eq:prop_const}.
Then, they add supply and decrease borrow demand in such a way that they keep the interest rate high, even though they are decreasing borrow demand.
The excess yield earned by the strategic user comes at the expense of others in the pool, if we assume that they are passive and/or static lenders (which appears to be true empirically valid~\cite{dc_2022}).
In Figure~\ref{fig:pid_utilization}, we draw the utilization, supply, and demand curves that the strategic user causes the protocol to have during the attack.

\noindent Suppose that at time $s$ a strategic user notices that that $U_{s-1} < U^*$.
At time $s$, the strategic user adds $D^* + (D^* - D_{s-1})$ units of borrow to the protocol such that $U_s = 2U^*$. That is, the current supply is fixed at $S_s$. The strategic user designs a $D^*$ such that: $2D^*/S_s = U_s = 2U^*$. 
The user then slowly reduces their borrow demand at each time $s+i$ for $i \in [k]$.
In particular, the user we consider a linear stepwise reduction in demand so that $D_{s+i} - D_{s+i-1} = -\frac{1}{k}D^*$.
After the user sequentially reduces the demand back to $D^*$, they close their loan and realize a profit or loss. \\ \\
\noindent Now, we need to compute the supplied assets that are needed for the strategic user to keep the utilization constant (and hence, the interest rate high).
At time $t=s+1$, the strategic user removes demand $\frac{1}{k} (2D^*-D_s)$ (leaving the user's net demand at $\left(1-\frac{1}{k}\right)(2D^* - D_{s-1})$) while adding supply $\Delta S_{s}$ so that $S_{s+1} = S_s + \Delta S_s$.
The amount of supply added needs to keep the utilization constant, e.g.
\[
U_{s+1} = \frac{\left(1-\frac{1}{k}\right)(2D^* - D_{s-1}) + D_{s-1}}{S_{s} + \Delta S_{s}} = \frac{\left(1-\frac{1}{k}\right)(2D^*) + \frac{1}{k}D_{s-1}}{S_s+\Delta S_s} = U^*
\]
which implies that
\[
\Delta S_{s} = S_s - \frac{2}{k}S_s + \frac{1}{k}\frac{D_{s-1}}{U^*}
\]
where we use $\frac{D^*}{U^*} = S_s$.
This implies that $S_{s+1} = S_s + \Delta S_s = 2S_s - \frac{2}{k}S_s + \frac{1}{k}\frac{D_{s}}{U^*}$.
We similarly can calculate $\Delta S_{s+1}$ via the equation
\[
U_{s+2} = \frac{\left(1-\frac{2}{k}\right)(2D^*) + \frac{2}{k}D_{s-1}}{S_{s+1} + \Delta S_{s+1}} = U^*
\]
which yields $\Delta S_{s+1} = \frac{1}{k}\left(\frac{D_{s-1}}{U^*} - 2S_s\right)$.
Using induction one can then show that $\Delta S_{s+i} = \frac{1}{k}\left(\frac{D_{s-1}}{U^*} - 2S_s\right)$ so that
\begin{align*}
    S_{s+i} &= S_s + \Delta S_{s} + \sum_{j=1}^i \Delta S_{s+j} = 2S_s + \frac{i+1}{k}\left(\frac{D_{s-1}}{U^*} - 2S_s\right) \\
    &= 2S_s \left(1 - \frac{i+1}{k}\right) + \frac{i+1}{k}\frac{D_{s-1}}{U^*}
\end{align*}
In particular, one can see that the amount of supply added at every timestep during the attack, $\Delta S_{s+i}$, is constant for all $i$. Furthermore, note that $D_{s-1} < D^*$ so that $\Delta S_{s+i} < -\frac{1}{k}S_s$ as $D^* = U^* S_s$.
This implies that
\[
S_{s+i} = 2S_s \left(1 - \frac{i+1}{k}\right) + \frac{i+1}{k}\frac{D_{s-1}}{U^*} < 2S_s \left(1 - \frac{i+1}{k}\right) + \frac{i+1}{k}S_s = \left(2 - \frac{i+1}{k}\right)S_s
\]
\noindent Using these demand and supply schedules, we can compute the cost and revenue inured by the strategic user.
Note that since we kept the utilization at $U^*$, the interest rate is constant throughout, \ie~$r_{s+i}(U^*) = r_s(U^*)$.
The cost $C_k$ for iterating this over $k$ blocks:
\begin{align*}
C_k &= r_s(U^*) \left(\sum_{i=1}^k \left(1-\frac{i}{k}\right) (2D^* - D_{s-1}) + (2D^*-D_{s-1})\right) \\
&= r_s(U^*)(2D^*-D_{s-1})\frac{k+1}{2}
\end{align*}
On the other hand, the revenue for performing this over $k$ blocks is
\begin{align}\label{eq:prop_revenue}
R_k &= r^S_s(U^*) \sum_{i=1}^k S_{s+i} = r^S_s(U^*) \left((k-3) S_s + \frac{k-3}{2} \frac{D_{s-1}}{U^*}\right) \nonumber \\
&=\gamma r_s(U^*) U^{*} (k-3)  \left(S_s + \frac{D_{s-1}}{2U^*}\right)
\end{align}
because $U_t = U^*$ is constant over the period of the entire attack. This gives us the profit over $k$ blocks, $P_k$, of
\begin{align*}
P_k &= \gamma r_s(U^*) U^* (k-3)  \left(S_s + \frac{D_{s-1}}{2U^*}\right) - r_s(U^*)(2D^*-D_{s-1})\frac{k+1}{2} \\
&= r_s(U^*)\left(k\underbrace{\left(\gamma \left(U^* S_s + \frac{D_{s-1}}{2}\right) - \frac{2D^*-D_{s-1}}{2}\right)}_{\text{duration dependent}}  - \overbrace{3 \left(U^* S_s + \frac{D_{s-1}}{2}\right) - \frac{2D^*-D_{s-1}}{2}}^{\text{hurdle rate}}\right)
\end{align*}
We highlight the two terms in the profit --- a duration dependent term (\eg~depends on the number of blocks $k$ that one holds a position for) and a `hurdle rate' or a constant profit that needs to be achieved to breakeven. \\ \\
\noindent The strategic user is profitable if two conditions hold:
\begin{enumerate}
    \item The duration dependent term is positive (so that the profit is $\Omega(k)$)
    \item $k$ times the hurdle rate is greater than the breakeven point
\end{enumerate}
Mathematically, the first term corresponds to the condition
\[
\gamma > \frac{2D^*-D_{s-1}}{2U^* S_s + D_{s-1}} =  \frac{2D^*-D_{s-1}}{2 D^* + D_{s-1}}
\]
because $U^* S_s = D^*$, whereas the latter condition corresponds to
\[
k > \frac{3}{\gamma}
\]
These conditions demonstrate that unless the reserve factor $\gamma$ is chosen in a sufficiently optimal manner, this attack can be successful. To interpret the first condition, if $D_{s-1} = 0$, then the above attack is \emph{always} unprofitable as the condition reduces to:
\begin{align*}
    \gamma > \frac{D^*}{D^*} = 1
\end{align*}
but this inequality is never achieved. This means that there must be sufficiently large latent demand present at time $s-1$ that the supplier can use to earn yield from. Alternatively, when $D_{s-1} = D^*$, we must have $\gamma > \frac{1}{3}$. That is, the spread must be sufficiently large for the supplier to be profitable. \\ \\ 
\noindent Notice that the spread/reserve factor $\gamma$ is a measure of capital efficiency (because the excess revenue goes to the protocol and not the lender) and as such reducing $\gamma$ reduces capital efficiency.
The above two conditions show that we effectively have to reduce $\gamma$ to lower attack profitability, illustrating the trade-off between this attack's success and capital efficiency.

\paragraph{New attacker.} This attack assumes that the attacker has no prior lending supply or demand in the protocol. An existing supply position would increase profitability of the attack and an existing demand position would decrease it. A weaker condition for profit extraction would be to consider an attack initiated by the largest existing net supplier, which we omit to show that P-controlled interest rates are vulnerable even when lending supply is highly distributed.

\subsection{Attack Mitigations}
\paragraph{Adjusting $k_t$ to elasticities.}
The attack as described above makes the worst case elasticity assumption, which is that $S_s, D_s$ do not change (outside of the strategic user's added supply and demand) in response to changes in interest rate.
In particular, if the borrow and supply series $S_s, D_s$ are stochastic processes, then we need to look at whether the strategic user has a positive \emph{expected profit}.
To do this, we can use simple heuristics based on concentration inequalities. \\ \\
\noindent More formally, define the expected supply and demand elasticities, $\Gamma^S, \Gamma^D$, as
\begin{align*}
\Gamma^S = \Expect\left[\frac{\partial S_s}{\partial r_s}\right] && \Gamma^D = \Expect\left[\frac{\partial D_s}{\partial r_s}\right]
\end{align*}
where the expectation is over the supply and demand time series $S_s, D_s$.
Note that in an individually rational rates market~\cite{bilson1978rational}, $\partial_{r_s}S_s \geq 0$ and $\partial_{r_s} D_s\leq 0$ so $\Gamma^S \geq 0, \Gamma^D \leq 0$. \\ \\
\noindent If the supply and demand elasticities are high, then the assumptions used in the prior section are less likely to hold, as Markov's inequality implies that
\begin{align*}
\Prob\left[\frac{\partial S_s}{\partial r_s} > \epsilon \right] < \frac{\Gamma^S}{\epsilon} && \Prob\left[\frac{\partial D_s}{\partial r_s} > \epsilon \right] < \frac{\Gamma^D}{\epsilon}
\end{align*}
These bounds allow for one to show that expected profitability $\Expect[P_k]$ (where the randomness comes from the other users generating supply and borrow) satisfies
\begin{align}\label{eq:ex_profit}
\Expect[P_k] &\geq  \gamma r_s(U^*) U^* (k-3)  \left(S_s + \frac{D_{s-1}}{2U^*}\right) - r_s(U^*)(2D^*-D_{s-1})\frac{k+1}{2}- \frac{Ck}{\epsilon}(\Gamma^S - \Gamma^D)
\end{align}
for a constant $C >0$.
We note that if other assumptions can be made (\eg~bounded increments), then stronger bounds can be computed using Azuma's inequality.
This can still be positive, especially when the elasticities $\Gamma^S, \Gamma^D$ are sufficiently small.
To ensure that $\Expect[P_k]$ is negative, we need to adjust the constant $k_s(U^*)$ such that the first term in the right hand side of \eqref{eq:ex_profit} is smaller than $\frac{Ck}{\epsilon}(\Gamma^S - \Gamma^D)$.

\paragraph{Dynamically Adjusting Spreads.}
A very clear sufficient condition for the strategic user to not be profitable is $\gamma \leq \frac{2D^*-D_{s-1}}{2\left(S_s + \frac{D_{s-1}}{U^*}\right)}$.
Recall that $\gamma$ is the spread between borrow and supply interest rates and if $\gamma = 1, r^S_t = r_t$.
The condition $\gamma \leq \frac{2D^*-D_{s-1}}{2\left(S_s + \frac{D_{s-1}}{U^*}\right)} < U^*$ is feasible when $U^*$ is near $1$ (\eg~target utilization is maxmimized).
However for many markets, the target utilization is usually quite far from $U^*$ for risk reasons. \\ \\
\noindent In particular, markets with predominantly recursive borrow such as the stETH/ETH market, have had trouble when utilization was near its maximum~\cite{morrow_2022}.
For these markets, one needs a target utilization $U^*$ that is closer to 50\% to discourage recursive borrowing.
However the condition $\gamma < U^*$ for $U^* = 0.5$ implies that the protocol is taking 50\% of earned revenue instead of giving it to lenders --- something that will cause elasticity to drop and make the attack more profitable. \\ \\ 
\noindent Newer lending protocols, such as Compound V3, allow for supply and demand rates to be adjusted separately, which implicitly means that $\gamma$ can be adjusted as a function of time.
In particular, a lending protocol could adjust $\gamma_t$ such that it maximizes both $\Gamma^S$ and $\frac{2D^*-D_{s-1}}{2\left(S_s + \frac{D_{s-1}}{U^*}\right)}$.
Such an estimate can be done off-chain using simulation (as the elasticities will need to be modeled) or on-chain using time-weighted elasticity measurements.

\paragraph{Protocol-Owned Liquidity.}
The protocol's treasury could be used to thwart such attack by adding in excess supply whenever there is an interest rate shock.
This is equivalent mathematically to having dynamically adjusted spreads, except that the protocol provides the funds instead of users incentivized by the change in spread.

\paragraph{Supply Caps}
One other strategy that can be used to thwart this attack is to not allow more than a certain amount of new supply to be added to the protocol if a single borrow causes the protocol to increase the rate dramatically.
These supply caps would need to be dynamic and adjusted to how fast the interest rate has changed after the addition of an individual loan.
Such supply caps would reduce the revenue term in the calculation of the previous section while also worsening user UX, especially for users who utilize the protocol via automatic rebalancing vaults such as Yearn Finance.

\paragraph{Capital Efficiency of Mitigations.}
All of these strategies for mitigating this attack are capital inefficient.
They either inure extract realized or opportunity costs on suppliers and borrowers.
However, not thwarting this attack effectively allows for those with large capital bases and infrastructure to actively trade to collect significantly more yield on their capital than passive borrowers and lenders.
Note that this attack can be much more profitable than Just-in-Time liquidity attacks~\cite{wan_adams_2022} as the yields in lending protocols tend to be significantly higher than those of automated market makers.

\section{Conclusion}
In this note, we formalized proportional and proportional-integral interest rate controllers in DeFi.
Using this formalism, we were able to construct an attack that a strategic user could use to collect excess yield from non-strategic passive users.
This formalism also allowed us to construct sufficient conditions for strategic user profitability and to define mitigating strategies.
These mitigations reduce the capital efficiency of the protocol, especially compared to static interest rate protocols. \\ \\ 
\noindent However, we were able to demonstrate that optimization of interest rate curves depends on supply and demand elasticities.
These elasticities can be measured and fed in as an input to future interest rate controllers.
Moreover, we demonstrate that properly tuned time-weighted averaging of interest rates can decrease attacker profitability in Appendix~\ref{app:PI_profit}.
Future work includes constructing a controller that utilizes elasticities in a manner that mitigates attacks of the form presented in this note.

\printbibliography
\appendix
\section{PI Controller Attack Profitability}\label{app:PI_profit}
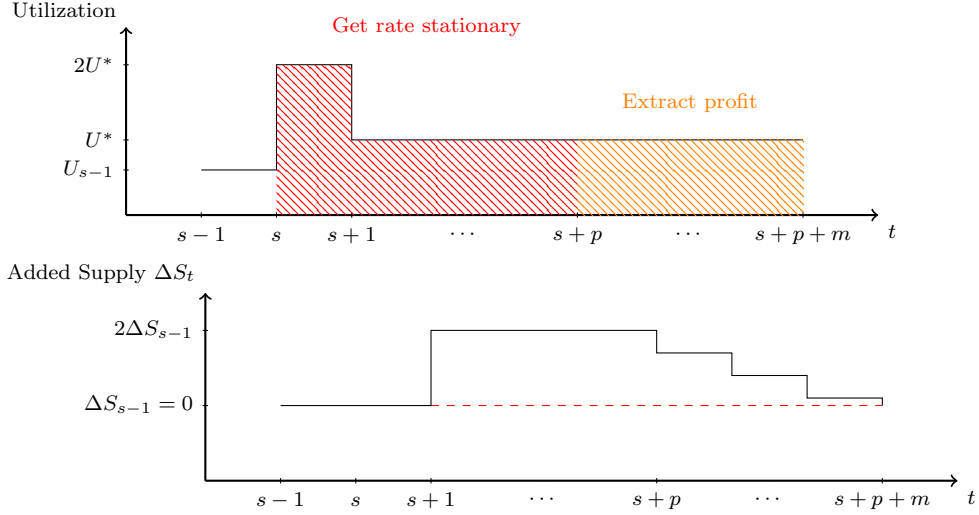
\begin{figure}
    \centering
    
    \begin{tikzpicture}
      \tikzstyle{every node}=[font=\footnotesize]

      \draw[thick,->] (0,0) -- (10,0) node[anchor=north west] {$t$};
      \draw[thick,->] (0,0) -- (0,2.5) node[anchor=south east] {Utilization};
      
      \draw (1 cm,1pt) -- (1 cm,-1pt) node[anchor=north] {$s-1$};
      \draw (2 cm,1pt) -- (2 cm,-1pt) node[shift={(0,-0.06)}, anchor=north] {$s$};
      \draw (3 cm,1pt) -- (3 cm,-1pt) node[anchor=north] {$s+1$};
      \draw (4.5 cm,0) -- (4.5 cm,0) node[shift={(0,-0.06)}, anchor=north] {$\cdots$};
      \draw (6 cm,1pt) -- (6 cm,-1pt) node[anchor=north] {$s+p$};
      \draw (7.5 cm,0) -- (7.5 cm,0) node[shift={(0,-0.06)}, anchor=north] {$\cdots$};
      \draw (9 cm,1pt) -- (9 cm,-1pt) node[anchor=north] {$s+p+m$};
      
      \draw (1pt,0.6 cm) -- (-1pt,0.6 cm) node[anchor=east] {$U_{s-1}$};
      \draw (1pt,1 cm) -- (-1pt,1 cm) node[anchor=east] {$U^*$};
      \draw (1pt,2 cm) -- (-1pt,2 cm) node[anchor=east] {$2U^*$};
      
      \draw (1,0.6) -- (2,0.6) -- (2,2) -- (3,2) -- (3,1) -- (4,1) -- (5,1) -- (6,1) -- (7,1) -- (8,1) -- (9,1);
      
      \path [pattern color=red, pattern=north west lines] (2,0) -- (6,0) -- (6,1) -- (3,1) -- (3,2) -- (2,2);
      \path [pattern color=orange, pattern=north west lines] (6,0) -- (9,0) -- (9,1) -- (6,1);
      \node [color=red, yshift=1.2cm] at (4,1.3) {Get rate stationary};
      \node [color=orange, yshift=1.2cm] at (7.5,0.3) {Extract profit};
    \end{tikzpicture}
    
    \begin{tikzpicture}
      \tikzstyle{every node}=[font=\footnotesize]

      \draw[thick,->] (0,0) -- (10,0) node[anchor=north west] {$t$};
      \draw[thick,->] (0,0) -- (0,2.5) node[anchor=south east] {Added Supply $\Delta S_t$};
      
      \draw (1 cm,1pt) -- (1 cm,-1pt) node[anchor=north] {$s-1$};
      \draw (2 cm,1pt) -- (2 cm,-1pt) node[shift={(0,-0.06)}, anchor=north] {$s$};
      \draw (3 cm,1pt) -- (3 cm,-1pt) node[anchor=north] {$s+1$};
      \draw (4.5 cm,0) -- (4.5 cm,0) node[shift={(0,-0.06)}, anchor=north] {$\cdots$};
      \draw (6 cm,1pt) -- (6 cm,-1pt) node[anchor=north] {$s+p$};
      \draw (7.5 cm,0) -- (7.5 cm,0) node[shift={(0,-0.06)}, anchor=north] {$\cdots$};
      \draw (9 cm,1pt) -- (9 cm,-1pt) node[anchor=north] {$s+p+m$};
      
      \draw (1pt,1 cm) -- (-1pt,1 cm) node[anchor=east] {$\Delta S_{s-1} = 0$};
      \draw (1pt,2 cm) -- (-1pt,2 cm) node[anchor=east] {$2 \Delta S_{s-1}$};
      
      \draw[dashed, color=red] (3,1) -- (9,1) node[right]{};
      \draw (1,1) -- (2,1) -- (3,1) -- (3,2) -- (6,2) -- (6,1.7) -- (7,1.7) -- (7,1.4) -- (8,1.4) -- (8,1.1) -- (9,1.1) -- (9, 1);
    \end{tikzpicture}
    
    \caption{Diagram of $U_t, \Delta S_t$ for the strategic user. The user initially spikes the interest rate at time $t$ and keeps the supply spiked until the interest rate is stationary (which takes $p$ time steps). Afterwards, they gradually lower their demand and supply until their demand was removed at time $s+p+m$.}
    \label{fig:pi_utilization}
\end{figure}

We will analyze the proportional-integral controller of equation~\ref{eq:prop_integral} with $\xi = 1$ and $\beta = \frac{\beta'(U_t)}{p}$.
Our attack will keep the utilization at $2U^*$ for 1 time step, then decrease the utilization to $U^*$ over $m$ time steps using a similar staircase function.
As illustrated in Figure~\ref{fig:pi_utilization}, the utilization is kept at $U^*$ long enough for the interest rate to stabilize before we gradually lower the supply committed.
For simplicity, we will against assume the elasticity is zero so that $U_{s-p} = U_{s-p+1} = \cdots = U_{s-1}$ (and the same for supply and demand) and that the original demand $D_{s-1} = D^*$.
All formulas can be modified to handle the case where this isn't true (albeit with signficantly more messy formulas).

Using the \eqref{eq:prop_integral} we have the following for $j \in [p]$
\begin{align}
    k_s(2U^*) &= k_{s-1}(U_{s-1}) + \alpha U^* + \beta \sum_{i=s-p}^{s-1} (U_i - U^*) = k_{s-1}(U_{s-1}) + \alpha U^* + \beta p (U_{s-1}-U^*) \nonumber \\
    k_{s+1}(U^*) &= k_{s}(2U^*) + \beta \sum_{i=s-p+1}^{s+1} (U_i - U^*) = k_s(2U^*) + \beta \left(U^* + (p-1)(U_{s-1}-U^*)\right) \nonumber \\ 
    k_{s+j}(U^*) &= k_{s}(2U^*) + \beta \sum_{i=s-p+j}^{s+j} (U_i - U^*) = k_s(2U^*) + \beta \left(U^* + (p-j)(U_{s-1}-U^*)\right) \label{eq:pi_recursion}
\end{align}
Note that $k_{s+p}(U^*) = k_s(2U^*) + \beta U^*$ and that $k_{s+p+i}(U^*) = K_{s+p}(U^*)$ for the schedule of Figure~\ref{fig:pi_utilization}. 
If we define $C^{stab}_p$ to be the cost for stabilizing the rate, we can write it as
\begin{align*}
    C^{stab}_p &= r_s(2U^*)(2D^*-D_{s-1}) + \sum_{i=1}^p r_{s+i}(U^*)(2D^*-D_{s-1}) \\
    &= U^*(2D^*-D_{s-1})\left(2k_S(2U^*) + \sum_{i=1}^p k_{s+i}(U^*)\right)
\end{align*}
Using equation~\eqref{eq:pi_recursion}, we can compute the sum in the above as follows
\begin{align}
    \sum_{i=1}^p k_{s+i}(U^*) &= k_s(2U^*) + \beta \sum_{i=1}^p U^* + (p-j)(U_{s-1}-U^*) \nonumber \\ 
    &= k_s(2U^*) + \beta\left(pU^* + p^2(U_{s-1}-U^*) - (U_{s-1}-U^*)\frac{p(p+1)}{2}\right)  \nonumber \\
    &= k_s(2U^*) + \beta p\left(U^* + \frac{p-1}{2}(U_{s-1}-U^*) \right) \label{eq:rate_sum}
\end{align}
which yields the final cost
\begin{equation}\label{eq:stab_cost}
C^{stab}_p = U^*(2D^*-D_{s-1})\left(3k_S(2U^*) +  \frac{\beta p}{2}\left((p-1)U_{s-1}-(p-3)U^*) \right)\right)
\end{equation}

Note that in Figure~\ref{fig:pi_utilization}, we show that the supply increases one step after the utilization increases.
This is to bring the utilization back to $U^*$, so we add supply at time $S_{s+1}$ until $S_{s+p}$.
The amount of supply we add needs to keep the utilization constant and solving for $\Delta S_s$ in
\[
U^* = \frac{2D^*}{S_s + \Delta S_s}
\]
yields $\Delta S_s = S_s$.
Now for each time $s+p+k$, we have a similar equation
\[
U^* = \frac{\left(2-\frac{k}{m}\right)D^*}{S_{s+p+k} + \Delta S_{s+p+k}}
\]
which leads to the same recursion as we found for the proportional controller, \ie~$\Delta S_{s+p+k} = -\frac{1}{m}S_{s+p}$.
Since, by construction, the interest rate is stationary at time $s+p+k$, we can compute the cost during the time period in orange in Figure~\ref{fig:pi_utilization} as
\[
C^{extract}_m = r_{s+p+1}(U^*) \sum_{i=1}^m \left(2 - \frac{i}{m}\right)D^* = \frac{3m-1}{2} D^* r_{s+p+1}(U^*) 
\]

Next, we need to compute our revenue.
We supply $S_s$ units between times $s+1$ and $s+p$ which earns interest $\gamma r_{s+i}(U^*) = \gamma k_{s+i}(U^*) U^*$.
Thus our initial supply revenue $R^{supp}_p$
\begin{equation}\label{eq:rsupp_p}
R^{supp}_p = \gamma U^* \sum_{i=1}^p k_{s+i}(U^*) = \gamma U^* \left(k_s(2U^*) + \beta p\left(U^* + \frac{p-1}{2}(U_{s-1}-U^*) \right)\right)
\end{equation}
where we used eq.~\eqref{eq:rate_sum}.
Our revenue between times $s+p$ and $s+p+m$ is the same as eq.~\eqref{eq:prop_revenue} for $m$ time steps.
Putting this all together, we receive a profit $P_{p,m}$ of
\begin{align*}
    P_{p, m} = R^{supp}_p + R_m - C^{stab}_p - C^{extract}_m
\end{align*}
These equations demonstrate that the cost is significantly higher than for the proportional controller, which can be seen from looking at the excess profit or loss $R^{supp}_p - C^{stab}_p$.
This cost is far more likely to be negative, especially since the cost has a term $3k_s(2U^*)$ versus the revenue's $\gamma k_s(2U^*)$.
We note, however, that there exist more complex strategies for PI controllers (e.g. adding and removing demand at different intervals to boost the rate earned during the stabilization period) than proportional controllers.


\end{document}